\documentclass[prb,twocolumn,showpacs,aps,superscriptaddress,floatfix]{revtex4-2}
\usepackage{amsmath}
\usepackage{amssymb}
\usepackage{bm}
\usepackage{graphicx}
\usepackage{color}
\usepackage{hyperref}
\usepackage{verbatim}

\begin{document}

\title{Spontaneous strain and magnetization in doped topological insulators with nematic and chiral superconductivity}

\author{R.S. Akzyanov}
\affiliation{Dukhov Research Institute of Automatics, Moscow, 127055 Russia}
\affiliation{Moscow Institute of Physics and Technology, Dolgoprudny,
    Moscow Region, 141700 Russia}
\affiliation{Institute for Theoretical and Applied Electrodynamics, Russian
    Academy of Sciences, Moscow, 125412 Russia}

\author{A.V. Kapranov}
\affiliation{Moscow Institute of Physics and Technology, Dolgoprudny,
    Moscow Region, 141700 Russia}

\author{A.L. Rakhmanov}
\affiliation{Dukhov Research Institute of Automatics, Moscow, 127055 Russia}
\affiliation{Institute for Theoretical and Applied Electrodynamics, Russian
    Academy of Sciences, Moscow, 125412 Russia}
\affiliation{Moscow Institute of Physics and Technology, Dolgoprudny,
    Moscow Region, 141700 Russia}

\begin{abstract}
We show that spontaneous strain and magnetization can arise in the doped topological insulators with a two-component superconducting vector order parameter. The details of the effects depend on the symmetry of the order parameter, whether it is nematic or chiral. The transition from the nematic state to the chiral one can be performed by application of a magnetic field, while the transition from the chiral state to the nematic is tuned by the external strain. These transitions associated with a jump of the magnetic susceptibility and mechanical stiffness. Possible experimental observations of the predicted effects are discussed.
\end{abstract}



\maketitle

\section{Introduction}

Topological superconductivity in doped topological insulators, such as A$_x$Bi$_2$Se$_3$, where A=Cu, Nb, Sr, possess unusual properties~\cite{Yonezawa2019}. Observation of the Knight shift gave an indication that the Cooper pairing here is triplet~\cite{Matano2016}. However, recent experiments show that this result requires careful verification and the statement about triplet nature of the pairing could be controversial~\cite{Pustogow2019,Ishida2020}. Measurements of the magnetoresistance show a two-fold symmetry of the second critical field $H_{c2}$ despite of three-fold crystal symmetry~\cite{Pan2016,Kuntsevich2018,Kuntsevich2019}. These properties are described in the framework of the theory of nematic superconductivity~\cite{Fu2010,Venderbos2016}. The nematic superconductor has a two-component order parameter with $E_u$ symmetry, which can be presented as a real-valued vector. The nematic superconductivity could be accompanied by such intriguing properties as surface Andreev bound states~\cite{Hao2017}, vestigial order~\cite{Hecker2018}, unconventional Higgs modes~\cite{Uematsu2019}, and Majorana fermions~\cite{Wu2017}. STM measuremts show a full gap in the spectrum~\cite{Levy2013,Wang2013} that can be attributed to the s-wave order parameter or to the effect of the hexagonal warping in the system with nematic superconducting order~\cite{Fu2014}. 

An alternative superconducting state with $E_u$ symmetry is a chiral phase~\cite{Fu2014}. In this phase, the time-reversal symmetry is broken and the order parameter is a two-component complex-valued vector. The chiral phase is predicted in thin films of doped topological insulators~\cite{PhysRevB.98.014505}. Recent experiments show fingerprints that can be attributed to the existence of the chiral superconductivity in the topological insulators~\cite{Huang2018,Kawai2020}.

One of the distinct features of the nematic superconductivity is a non-trivial coupling with strain. This coupling leads to the two-fold symmetry of $H_{c2}$~\cite{Venderbos2016}. In Refs.~\onlinecite{Kuntsevich2018,Kuntsevich2019} this effect was observed. In Ref.\cite{Kuntsevich2019} it was found that strain also affects two-fold symmetric anisotropic magnetoresistance and breaks three-fold crystal symmetry. The X-ray studies reveal that in most samples initial strain about $\delta l/l \sim10^{-5}$ is presented at room temperature. In some samples, the deformation at room temperature is absent up to the experimental accuracy, while a two-fold symmetry of $H_{c2}$ remains. In Ref.~\onlinecite{Cho2020} experiment on the magnetostriction reveals that the crystal lattice is deformed in the superconducting state. The strain is $\delta l/l \sim10^{-7}$ and decreases with an increase of temperature and vanishes in the normal state. These experiments demonstrate that either spontaneous or initial strain probably exists in the doped topological insulators in the superconducting state. 

In Ref.~\onlinecite{arxiv:1512.03519} the existence of a magnetization in the superconducting state of Nb$_{x}$Bi$_2$Se$_3$ has been found. The magnetization vanishes in the normal state. The origin of such magnetization is yet to be clarified. DFT calculations show that the magnetization can be attributed to the existence of the free spins in the intercalated Nb adatoms. However, the spontaneous magnetization has not been observed in subsequent experiments and it is debated whether it actually exists~\cite{Yonezawa2019}. Recently, a non-zero magnetization in the superconducting Sr$_{0.1}$Bi$_2$Se$_3$ have been observed using muon spectroscopy~\cite{Neha2019}.

In Refs.~\onlinecite{Venderbos2016} and \onlinecite{Fu2014}, the Ginzburg-Landau (GL) functional was derived for the superconducting state in the topological insulator with vector order parameter $\vec{\eta}=(\eta_1,\eta_2)$. The nematic state corresponds to real order parameter $\vec{\eta}=\eta(\cos \alpha, \sin \alpha)$, while in the chiral state the order parameter is a complex vector $\vec{\eta}=\eta(1, \pm i)$. The GL theory successfully describes macroscopic properties of the superconductivity in the doped topological insulators. 

The multicomponent structure of the order parameter allows emergence of the subsidiary order parameters. In the case of $D_{3d}$ point group symmetry of topological insulators, such as Bi$_2$Se$_3$, the subsidiary point groups have $E_g$ and $A_{2g}$ symmetries. Corresponding bilinear forms are~\cite{Venderbos2016}
\begin{eqnarray*}
E_{g} &\rightarrow & (N_1,N_2)=(|\eta_1|^2-|\eta_2|^2,\eta_1^*\eta_2+\eta_1\eta_2^*),\\
A_{2g} &\rightarrow & M_0=\eta_1\eta_2^*-\eta_1^*\eta_2.
\end{eqnarray*}
As we can see, the nematic state corresponds to $E_g$ subsidiary order with $M_0=0$, while the chiral state corresponds to $A_{2g}$ bilinear with $(N_1,N_2)=(0,0)$. It was pointed out in Ref.~\onlinecite{Venderbos2016} that the existence of the subsidiary order parameters gives rise to a non-trivial coupling of the superconductivity with magnetization and strain. Corresponding contribution to the GL free energy has a structure $2iM_z M_0+(u_{xx}-u_{yy})N_1+2u_{xy}N_2$. Thus, the nematic state couples with the strain degrees of freedom, while chiral state couples with the magnetization.  In other words, nematicity and deformation competes with the chirality and magnetization. While we consider theory for the specific symmetry group, we expect similar effects in other materials with appropriate symmetry of the superconducting order.

Here, we show that either spontaneous strain or spontaneous transverse magnetization arises in the topological superconductor depending on the system parameters. The spontaneous deformation is observed in the case of nematic symmetry of the superconducting order parameter, $\vec{\eta}=\eta(\cos \alpha,\sin \alpha)$, while the spontaneous magnetization exists in the case of the chiral solution, $\vec{\eta}=\eta(1,\pm i)$. We also study the effects of initial strain and applied magnetic field on the order parameter symmetry. We show that the growth of the applied magnetic field gives rise to a transition of the nematic order to the chiral one, while with the growth of the initial strain, the chiral state transits to the nematic one. We found that the magnetic susceptibility and the stiffness experience jumps at the transition points. We discuss the relation of the obtained results with recent experimental observations. 


\section{GL free energy}\label{GL_f}

We study the superconducting state in the doped topological insulator assuming that the system is spatially uniform and the order parameter is independent of coordinates. In particular, we do not consider Abrikosov vortices. In presence of the half-quantum vortices with topological charge\cite{Zyuzin2017} our theory is not applicable and more advanced theory of deconfined quantum criticality should be applied\cite{Senthil2004}. Uniform GL free energy can be written in the form~\cite{Fu2014,Venderbos2016}
\begin{eqnarray}
F_0&=&A(|\eta_1|^2+|\eta_2|^2)+B_1(|\eta_1|^2+|\eta_2|^2)^2\\
&+&B_2|\eta_1^*\eta_2-\eta_1\eta_2^*|^2,
\end{eqnarray}
where GL coefficients $A\propto T-T_c<0$, $B_1>0$, and $B_2$ either positive or negative. Minimization of this GL free energy predicts the existence of two different superconducting states depending on the sign of $B_2$:
\begin{eqnarray}\label{States1}
\!\!\!\!\!\!\!\!\!&&|\eta_1|^2\!\!+\!\!|\eta_2|^2\!=\!-\frac{A}{2B_1},\, \textrm{Im}(\eta_1,\eta_2)\!=\!0,\, F_0\!=\!-\frac{A^2}{4B_1},\, B_2>0,\nonumber\\
\!\!\!\!\!\!\!\!\!&&|\eta_1|^2\!\!=\!\!\frac{-A}{4(B_1\!\!+\!\!B_2)},\,\, \eta_1\!\!=\!\!i\eta_2,\,\, F_0\!\!=\!\!-\frac{A^2}{4(B_1\!\!+\!\!B_2)},\,\, B_2\!<\!0.
\end{eqnarray}
The state with $B_2>0$ is referred to as nematic. This state has a real order parameter $\vec{\eta}=\eta(\cos \alpha,\sin \alpha)$. The state with $B_2<0$ is commonly called chiral. In the chiral state, the time-reversal symmetry is broken, which is related to the complex order parameter $\vec{\eta}=\eta(1,\pm i)$.

The total GL free energy is a sum   
\begin{equation}\label{GL0}
F_{\textrm{GL}}=F_0+F_u+F_M,
\end{equation}
where $F_0$ is the bare superconducting part given by Eq.~\eqref{States1}, $F_u$ and $F_M$ are the symmetry-breaking terms arising due to coupling of the superconducting order with the strain and the magnetization, respectively. 

The term $F_u$ for the system with $D_{3d}$ group symmetry can be presented as~\cite{Fu2014,Landau7}  
\begin{widetext}
\begin{eqnarray}\label{GL22}
F_u=g_N(u_{xx}-u_{yy})(|\eta_1|^2-|\eta_2|^2)+2g_Nu_{xy}(\eta_1^*\eta_2+\eta_1\eta_2^*)+\lambda_1\left[(u_{xx}-u_{yy})^2+4u_{xy}^2\right] + \lambda_2 (u_{xx}+u_{yy})^2,
\end{eqnarray}
\end{widetext}
where $u_{ik}$ are components of the strain tensor, $u_{xx}$ and $u_{yy}$ are uniaxial components, $u_{xy}=u_{yx}$ is a shear strain, and $g_N$ is a GL coupling constant between superconducting order and strain. Two last terms in Eq.~\eqref{GL22} corresponds to the self energy of the elastic deformation and $\lambda_1,\lambda_2>0$ are elastic modules. 

The term with the transverse Zeeman magnetization $M_z$ reads   
\begin{equation}\label{GLM0}
F_M=-2ig_MM_z(\eta_1\eta_2^*-\eta_1^*\eta_2)+aM_z^2,
\end{equation}
where $g_M$ is a GL coupling constant between superconducting order and the magnetization, and an empiric coefficient $a>0$. The last term in this equation accounts for the free energy loss due to magnetization in a non-magnetic phase. 

\section{Spontaneous strain and magnetization}\label{u_M}

We have to minimize the GL free energy with respect to $\vec{\eta}$, $u_{ik}$, and $M_z$. It is convenient to introduce following notations $\eta_1=\eta\cos \alpha\exp(i\varphi_1)$, $\eta_2=\eta\sin \alpha\exp(i\varphi_2)$, and $\varphi=\varphi_1-\varphi_2$. In these terms we have 
\begin{equation}\label{GL1}
F_0=A\eta^2+B_1\eta^4+B_2\eta^4\sin^2 2\alpha \sin^2 \varphi,
\end{equation}
\begin{equation}\label{GLM}
F_M=2g_M M_z \eta^2 \sin 2\alpha \sin \varphi+aM_z^2,
\end{equation}

We consider a uniform 2D strain in the system with hexagonal symmetry, which is characterized by three independent values $u_{xx}$, $u_{yy}$, and $u_{xy}$. Following Ref.~\onlinecite{How2019}, we use their linear combination as new independent variables. We divide the strain tensor $u_{ik}$ into two parts, a vector $\vec u= (u_{xx}-u_{yy},2u_{xy})=u(\cos2\beta,\sin2\beta)$ and a scalar $\text{Sp}\,u_{ik}=u_{xx}+u_{yy}$. In these notations Eq.~\eqref{GL22} is rewritten as  
\begin{eqnarray}\label{GL2}
\nonumber
F_u&=&g_N\eta^2 u (\cos 2\alpha \cos 2\beta+\cos \varphi \sin 2\alpha \sin 2\beta )\\
&+&\lambda_1 u^2+\lambda_2(u_{xx}+u_{yy})^2.
\end{eqnarray}

To calculate the spontaneous strain and transverse magnetization, we should minimize the total free energy with respect to new independent degrees of freedom: the magnitude of the order parameter $\eta$, the direction of the nematicity $\alpha$, the phase difference between components of the order parameter $\varphi$, the magnetization $M_z$, the amplitude of the strain $u$, the direction of $\vec{u}$, that is, angle $\beta$, and the trace of the deformation tensor $\text{Sp}\,u_{ik}$. The minimization by the latter variable is trivial: minimum of $F_{\textrm{GL}}$ attains if $\text{Sp}\,u_{ik}=0$ since this value does not couple with superconducting order. Thus, $u_{xx}=-u_{yy}$ and the spontaneous strain occurs without change of the sample volume.  

The minimization of $F_{GL}$ with respect to the magnetization means that $\partial F_M/\partial M_z=0$ and from Eq~\eqref{GLM} we derive
\begin{equation}\label{Mz}
M_z=-\frac{g_M}{a}\eta^2 \sin 2\alpha \sin{\varphi}.
\end{equation}
The spontaneous magnetization, which arises after normal to superconductor transition, exists only if $\sin 2\alpha$ and $\sin{\varphi}\neq 0$, that is, when both $\eta_1$ and $\eta_2$ are non-zero and the order parameter is complex.  

Minimization by other degrees of freedom is straightforward but cumbersome. We present only final results. We obtain that the system has two possible ground states, nematic and chiral, as in the case when we neglect the spontaneous strain and magnetization. In the nematic phase, $\sin \varphi=0$, we get
\begin{equation}\label{nem}
\eta^2=\frac{-A}{2(B_1\!-\!g_N^2/4\lambda_1)},\quad F_{\text{nem}}=\frac{-A^2}{4(B_1\!-\!g_N^2/4\lambda_1)}.   
\end{equation}
The condition $B_1-g_N^2/4\lambda_1>0$ is necessary for the stability of the nematic state. In the nematic state the spontaneous magnetization is zero, while the strain is non-zero and $u=u_{sp}(-1)^{n+1}$, where 
\begin{equation}\label{strain_nem}
u_{sp}=\frac{g_N |A|}{4B_1\lambda_1-g_N^2},\,\,\beta=\alpha + \frac{\pi n}2, \quad M_z=0,    
\end{equation}
and $n$ is integer. The deformation vector is parallel to the nematicity direction if $g_N>0$
\begin{equation}
\vec{u}=(u_{xx}-u_{yy},2u_{xy})=|u| (\cos 2\alpha , \sin 2\alpha)   
\end{equation}
and orthogonal if $g_N<0$. The nematic phase is infinitely degenerate with respect to the angle $\alpha$, and the ground state free energy is the same for any direction of the nematic order parameter $\vec{\eta}=\eta(\cos \alpha, \sin \alpha)$. The spontaneous deformation, $u_{sp} \propto A \propto (T_c-T)$, decreases with the increase of temperature and vanishes at $T=T_c$.  

In the chiral state $\cos \varphi=0$ and $\alpha=\pi/4+\pi l/2$, where $l$ is integer. In so doing, we derive
\begin{equation}\label{GL_ch}
\!\!\!\!\!\eta^2\!=\!\frac{-A}{2(B_1\!+\!B_2\!-\!g_M^2/a)},\,\,\, F_{\text{ch}}\!=\!\frac{-A^2}{4(B_1\!+\!B_2\!-\!g_M^2/a)}.   
\end{equation}
The system is stable if $B_1\!+\!B_2\!-\!g_M^2/a>0$. In the case of chiral phase, the spontaneous strain is zero, while the spontaneous magnetization is non-zero and $M_z=M_{sp}(-1)^{l+1}$, where
\begin{equation}\label{magn_ch}
M_{sp}=\frac{g_M |A|}{2\left[(B_1+B_2)a-g^2_M\right]}, \quad u=0.    
\end{equation}
The chiral state is degenerate with respect to the sign of chirality, $\eta=\eta(1,\pm i)$, and sign of the magnetization $M_z$. The spontaneous magnetization, $M_z \propto A \propto (T_c-T)$, decreases with the increase of temperature and vanishes at $T=T_c$.

We compare the free energy in the nematic, Eq.~\eqref{nem}, and in the chiral states, Eq.~\eqref{GL_ch}, and conclude that the nematic phase is the ground state, $F_{\textrm{nem}}<F_{\textrm{ch}}$, if 
\begin{equation}\label{cond_nem}
B_2+\frac{g_N^2}{4\lambda_1}-\frac{g_M^2}{a}>0.    
\end{equation}
Otherwise, the superconductor is in the chiral phase. 

As we can see from Eq.~\eqref{cond_nem}, coupling the superconductivity with the strain shifts the system toward the nematic state, while the coupling with the magnetization drives the system to the chiral state. Below, we show how we can switch off the system from one phase to another by application of a magnetic field or an external strain.

\section{Effect of the applied magnetic field} \label{field}

We assume that a uniform transverse magnetic field $\mathbf{H}=(0,0,H)$ exists in the sample volume. Such a situation could be realized, for example, if a corresponding size of the sample smaller than the London penetration depth. In this case, the field $H$ is simply an external magnetic field. Our consideration is valid if this field is much smaller than the upper critical field $H_{c2}$ when we can neglect the Landau quantization and, consequently, disregard spatial modulations of the order parameter. In other words, we consider an external Zeeman magnetization. We choose $z$-axis directed along the applied field and, hence, $H>0$. In this section, we neglect the spontaneous strain since it does not affect the main result but makes the calculations cumbersome.    

We rewrite Eq.~\eqref{GLM} for magnetic part of the GL free energy as 
\begin{equation}\label{GLM_H}
F_M=2g_M M_z \eta^2 \sin 2\alpha \sin \varphi+aM_z^2-HM_z.
\end{equation}
The minimization of $F_M$ with respect to $M_z$ gives
\begin{equation}\label{Mz_H}
M_z=\frac{1}{2a}\left(H-2g_M\eta^2 \sin 2\alpha \sin{\varphi}\right).
\end{equation}
After minimization of $F_{\textrm{GL}}$ by $\eta^2$ we obtain 
\begin{equation}
\eta^2=-\frac{A+g_MH \sin 2\alpha \sin \varphi/a}{2\left[B_1+(B_2-g_M^2/a)\sin^2 2\alpha \sin^2 \varphi\right]}.    
\end{equation}
We substitute expressions for $\eta^2$ and $M_z$ in Eqs.~\eqref{GL1} and \eqref{GLM_H} and derive    
\begin{equation}\label{Free_MH}
F_{\textrm{GL}}(t)=-\frac{H^2}{4a}-\frac{(A+g_MHt/a)^2}{4\left[B_1+(B_2-g^2_M/a)t^2\right]},    
\end{equation}
where $t=\sin 2\alpha \sin \varphi$ and $|t|\leq 1$. Edge values $t=\pm 1$ correspond to the chiral phase $\varphi = \pm \pi/2$, $\alpha = \pm \pi/4$, and the first minimum of the free energy is attained if $t=-1$. As a result, we have that in the chiral phase:   
\begin{equation}\label{Free_MH_ch}
F_{\textrm{ch}}(H)=-\frac{H^2}{4a}-\frac{(A-g_MH/a)^2}{4\left(B_1+B_2-g^2_M/a\right)}    
\end{equation}
and magnetization
\begin{eqnarray}\label{ch_MH}
M_z&=&\frac{H}{2a}\left[1+\frac{g_M^2}{a\left(B_1+B_2-g_M^2/a\right)}\right]+M_{sp}.
\end{eqnarray}
Here the first term is an induced magnetization and the second term is the spontaneous magnetization, Eq.~\eqref{magn_ch}.

We find the second minimum of the free energy from the condition $\partial F_{\textrm{GL}}(t)/\partial t$=0 at
\begin{equation}\label{tt}
t=\frac{g_MHB_1}{aA\left(B_2-g_M^2/a\right)}.    
\end{equation}
The latter solution exists only if $t<1$ or $H<|A|(aB_2-g_M^2)/g_MB_1$. This minimum corresponds to the nematic state and $\sin \varphi=0$ if $H=0$. In the nematic state, the order parameter is independent of $H$ and only induced magnetization is observed:
\begin{equation}\label{nem_eta_M}
\eta^2=-\frac{A}{2B_1},\qquad M_z=\frac{H}{2a}\left(1+\frac{g^2_M}{aB_2-g^2_M}\right).    
\end{equation}
The GL energy in the nematic phase is
\begin{equation}\label{Free_MH_nem}
F_{\textrm{nem}}(H)=-\frac{H^2}{4a}\left(1+\frac{g^2_M}{aB_2-g^2_M}\right)-\frac{A^2}{4B_1}.    
\end{equation}

If we assume that inequality~\eqref{cond_nem} is fulfilled, then, the nematic phase is the ground state at $H=0$. The applied magnetic field induces a non-zero phase difference $\sin \varphi = t/\sin 2\alpha$ between the components of the order parameter and drives the nematic state to the chiral one. Using Eq.~\eqref{tt} and comparing the GL free energies in Eqs.~\eqref{Free_MH_ch} and \eqref{Free_MH_nem}, we conclude that the value of $|t|$ increases with $H$ and attains its maximum $|t|=1$ at which $F_{\textrm{nem}}(H)
=F_{\textrm{ch}}(H)$ when  
\begin{equation}\label{threshold}
H=H^*=|A|(aB_2-g_M^2)/g_MB_1.    
\end{equation}
With further increase of the magnetic field, the nematic state disappears, the superconducting state becomes chiral. Correspondingly, a spontaneous magnetization arises at $H>H^*$. However, at this point the magnetization $M_Z(H)$ is continuous, while the magnetic susceptibility, $\chi=\partial M_z/\partial H$, exhibits a jump:  
\begin{eqnarray}\label{chi}
\chi_{\textrm{nem}}&=&\frac{1}{2a}\left(1+\frac{g^2_M}{aB_2-g^2_M}\right),\quad H<H^*,\\
\nonumber
\chi_{\textrm{ch}}&=&\frac{1}{2a}\left[1+\frac{g_M^2}{a\left(B_1+B_2-g_M^2/a\right)}\right],\quad H>H^*.
\end{eqnarray}
Thus, the transition from the nematic to chiral phase at $H=H^*$ is a type-II phase transition. 

\section{System with initial strain}\label{strain}

Here we assume that some initial strain $\vec{u}_0=u_0(\cos{2\beta_0},\sin{2\beta_0})$ exist in the system. According to X-ray measurements~\cite{Kuntsevich2018,Kuntsevich2019}, this strain arises in the process of the crystal growth and had a characteristic value $u_0\approx 10^{-5}$, which is two orders of magnitude larger than that observed after normal to superconductor transition~\cite{Cho2020}. Thus, it is reasonable to neglect here the spontaneous deformation. 

We have to minimize the free energy $F_{\textrm{GL}}=F_0+F_M+F_u(u_0)$, see Eqs.~\eqref{GL1}, \eqref{GLM}, and \eqref{GL2}. We neglect terms proportional to $u_0^2$ in $F_u$ since they are constant and, for definiteness, we assume that $g_Nu_0>0$. From the minimization condition  
$\partial F/\partial \eta =0$ and $\partial F/\partial M_z=0$, we obtain
\begin{eqnarray*}
\eta^2\!&=&\!-\frac{A\!+\!g_Nu_0(\cos 2\alpha \cos 2\beta_0\!+\!\cos \varphi \sin 2\alpha \sin 2\beta_0)}{2[B_1\!+\!(B_2\!-\!g_M^2/a)\sin^2 2\alpha \sin^2 \varphi]},\\   
F\!&=&\!-\frac{[A\!+\!g_Nu_0(\cos 2\alpha \cos 2\beta_0\!+\!\cos \varphi \sin 2\alpha \sin 2\beta_0)]^2}{4[B_1\!+\!(B_2\!-\!g_M^2/a)\sin^2 2\alpha \sin^2 \varphi]}.   
\end{eqnarray*}
The free energy $F$ has two minimums. The first of them corresponds to the nematic state, $\sin \varphi=0$. In this state the  spontaneous magnetization is absent and the free energy is
\begin{equation}\label{nem_u}
F_{nem}(u_0)=-\frac{\left[A+g_Nu_0 \cos 2(\alpha-\beta_0)\right]^2}{4B_1}.    
\end{equation}
In the ground state we have $\alpha=\beta_0+\pi(n+1/2)$ and $F_{\textrm{nem}}=-({A-g_Nu_0)^2/4B_1}$.

The second minimum of the free energy corresponds to the chiral state, where
\begin{eqnarray}\label{chi_angle}
\nonumber
\cos{2\alpha}&=&-\frac{u_0\cos{2\beta_0}}{u^*},\quad
\nonumber
\cos{\varphi}=-\frac{u_0\sin{2\beta_0}}{u^*\sin{2\alpha}},\\
u^*&=&\frac{|A|(g_M^2 - B_2a)}{g_N[(B_1+B_2)a-g_M^2]}.
\end{eqnarray}
The order parameter in the chiral state is the same as in the case $u_0=0$, Eq.~\eqref{GL_ch}. For the free energy and magnetization we have 
\begin{eqnarray}\label{chi_u0}
F_{ch}(u_0)&=&\frac{-A^2+g_N^2u_0^2(1+\frac{B_1}{B_2-g_M^2/a})}{4\left(B_1+B_2-g_M^2/a\right)},\\
\nonumber
M_z(u_0)&=&M_{sp}\sqrt{1-\left(\frac{u_0}{u^*}\right)^2}.
\end{eqnarray}

Comparing the free energies in Eqs.~\eqref{nem_u} and \eqref{chi_u0}, we see that the nematic phase is the ground state at any $u_0$, if the condition in Eq.~\eqref{cond_nem} is fulfilled. However, changing the the strain $\vec{u}_0=(u_{0xx}-u_{0yy},2u_{0xy})=u_0(\cos{2\beta_0},\sin{2\beta_0})$ we can govern the nematicity vector $\vec{\eta}=\eta(\cos{\alpha},\sin{\alpha})$ since $\alpha=\beta_0+\pi(n+1/2)$. For example, if the initial strain has no a shear component, that is, $\beta_0=0$, we have $\vec{\eta}=\eta(0,\pm 1)$. In the case of a pure shear strain, $\beta_0=\pi/4$ we get $\alpha=\pi/4+\pi(n+1/2)$ and $\vec{\eta}=\eta(1,\pm 1)/\sqrt{2}$. 

If the condition in Eq.~\eqref{cond_nem} is violated, then the ground state of the system at $u_0=0$ is chiral with non-zero spontaneous magnetization.  The applied strain drives the chiral phase to the nematic one via tuning of the $\alpha$ or $\varphi$ from the $\alpha=\pm\pi/4$ or $\varphi=0,\pi$, see Eq.~\ref{chi_angle}. If $u_0>u^*$ then Eq.~\ref{chi_angle} cannot be fulfilled and chiral phase becomes unstable and the ground state becomes nematic with zero $\varphi$ and $M_z$. Similar to the magnetic susceptibility, stiffness $K=\partial^2F/\partial u_0^2$ experience here a jump: 
\begin{eqnarray}\label{chik}
K_{\textrm{nem}}&=&2\lambda_1-\frac{g_N^2}{2B_1},\quad u>u^*,\\
\nonumber
K_{\textrm{ch}}&=&2\lambda_1+\frac{g_N^2}{2(B_2-g_M^2/a)},\quad u<u^*,
\end{eqnarray}
and the transition from the chiral to nematic state is of the second order.

\section{Discussion}\label{disco}

In the framework of the GL approach, we analyzed the symmetry breaking phenomena in the topological superconductors. We predict that in the nematic state a spontaneous strain of the crystal occurs due to a non-trivial coupling of the superconducting order parameter and the strain. This strain vanishes in the normal state. A spontaneous deformation has been observed in the measurements of the magnetostriction~\cite{Cho2020}: the lattice strain $\delta l/l=u \sim 10^{-7}$ arises in the superconducting state, decreases with  temperature growth, and vanishes in the normal state, which confirms our prediction. Similar values of the magnetostriction has been observed in other types of the superconductors~\cite{Eremenko1999}.

In Refs.~\onlinecite{Kuntsevich2018,Kuntsevich2019}, the strain has been measured by the X-ray technique. In the most samples, the initial deformation $u_0 \sim 10^{-5}$ at room temperature was found. It is much larger than the spontaneous deformation $u \sim 10^{-7}$ reported in Ref.~\onlinecite{Cho2020}. In some samples, no strain has been found up to the experimental accuracy. However, in all samples a two-fold in-plane anisotropy of $H_{c2}$ was observed. These experiments indicate that both cases, either dominant of the initial or spontaneous strain, are possible. Initial deformation $u_0$ increases the critical temperature $T_c=T_{c0}+|g_N u/A_0|$, where $T_{c0}$ is the critical temperature without strain, $A=A_0(T_{c0}-T)$. The spontaneous strain, $u \propto A_0(T_{c0}-T)$, vanishes at $T=T_{c0}$ and has no effect on the critical temperature. 

We argue that a transverse spontaneous magnetization occurs in the chiral state. This magnetization is tied to the superconductivity and vanishes in the normal state. The finite magnetization in the doped topological insulator Nb$_x$Bi$_2$Se$_3$ have been measured in Refs.~\onlinecite{arxiv:1512.03519,Neha2019}. This magnetization decreases with an increase in temperature and vanishes in the normal state, which is in agreement with our results. 

We consider the effects of the applied magnetic field and the initial strain. We show that the application of the magnetic field drives the initially nematic superconductor to the chiral state, while the growth of the initial strain derives the chiral superconductor to the nematic state. When the applied field (initial strain) exceeds some threshold value the nematic (chiral) state is changed by the chiral (nematic) phase. 

\section*{Acknowledgment}

RSA acknowledges the support by the Russian Scientific Foundation under Grant
no 20-72-00030 and partial support from the Foundation for the Advancement of Theoretical Physics and Mathematics “BASIS”.

\bibliography{bib}
\end{document}